\documentclass[twocolumn,showpacs,amsfonts,aps,prc,nofootinbib,floatfix]{revtex4}

\usepackage{bm}
\usepackage{graphicx}
\usepackage{amsmath}
\usepackage[usenames,dvipsnames]{color}
\usepackage{soul}

\begin{document}
\setstcolor{red}


\title{Thermal photons as a quark-gluon plasma thermometer reexamined}

\author{Chun Shen}
\email[Correspond to\ ]{shen@mps.ohio-state.edu}
\author{Ulrich Heinz}
\affiliation{Department of Physics, The Ohio State University,
  Columbus, OH 43210-1117, USA}
\author{Jean-Fran\c{c}ois Paquet}
\affiliation{Department of Physics, McGill University, 3600 University Street, Montreal, Quebec, H3A 2T8, Canada}
\author{Charles Gale}
\affiliation{Department of Physics, McGill University, 3600 University Street, Montreal, Quebec, H3A 2T8, Canada,}
\affiliation{Frankfurt Institute for Advanced Studies, Ruth-Moufang-Str. 1, D-60438 Frankfurt am Main, Germany}

\begin{abstract}
Photons are a penetrating probe of the hot medium formed in heavy-ion collisions, but they are emitted from all collision stages. At photon energies below 2-3 GeV, the measured photon spectra are approximately exponential and can be characterized by their inverse logarithmic slope, often called ``effective temperature'' $T_\mathrm{eff}$. Modelling the evolution of the radiating medium hydrodynamically, we analyze the factors controlling the value of $T_\mathrm{eff}$ and how it is related to the evolving true temperature $T$ of the fireball. We find that at RHIC and LHC energies most photons are emitted from fireball regions with $T{\,\sim\,}T_\mathrm{c}$ near the quark-hadron phase transition, but that their effective temperature is significantly enhanced by strong radial flow. Although a very hot, high pressure early collision stage is required for generating this radial flow, we demonstrate that the experimentally measured large effective photon temperatures $T_\mathrm{eff}{\,>\,}T_\mathrm{c}$, taken alone, do not prove that any electromagnetic radiation was actually emitted from regions with true temperatures well above $T_\mathrm{c}$. We explore tools that can help to provide additional evidence for the relative weight of photon emission from the early quark-gluon and late hadronic phases. We find that the recently measured centrality dependence of the total thermal photon yield requires a larger contribution from late emission than presently encoded in our hydrodynamic model.  
\end{abstract}


\date{\today}

\maketitle

Photons produced in heavy-ion collisions interact only electromagnetically and are thus able to penetrate the medium from which they are emitted without rescattering. Their usefulness for experimentally accessing the temperature of the quark-gluon plasma (QGP) created in ultra-relativistic nuclear collisions was first pointed out several decades ago \cite{Feinberg:1976ua,Shuryak:1978ij,Kajantie:1981wg}. The realization that strong collective flow generated during the expansion of the QGP will significantly affect the photon and dilepton transverse momentum ($p_T$) spectrum (but not the momentum-integrated invariant dilepton mass spectrum!) is almost as old \cite{Kajantie:1986cu}. Since photons are emitted from all stages of the collision, their momentum distributions integrate over the temperature and flow history of the expanding fireball, weighting it with emission rates that depend on the collision stage and the corresponding radiating degrees of freedom \cite{Turbide:2007mi}. The interpretation of the shape of experimentally measured photon spectra is therefore complex and requires theoretical modeling based on cross-checks with other experimental observables.

Recently the PHENIX and ALICE experiments measured an excess of direct photon production, attributed to thermal radiation, in 200\,$A$\,GeV Au+Au collisions at RHIC \cite{Adare:2008ab} and in 2.76\,$A$\,TeV Pb+Pb collisions at the LHC \cite{Wilde:2012wc}. In the low-$p_T$ region the direct photon spectra are approximately exponential and can be well characterized by their inverse logarithmic slope $T_\mathrm{eff}$:  $\frac{dN}{dy\,p_Tdp_T} \propto e^{-p_T/T_\mathrm{eff}}$. The PHENIX Collaboration reported $T_\mathrm{eff} = 221 \pm 19 \pm 19$\,MeV for Au+Au collisions with $0{-}20\%$ centrality at top RHIC energy \cite{Adare:2008ab} while the ALICE Collaboration found $T_\mathrm{eff} = 304 \pm 51^\mathrm{sys.+stat.}$\,MeV for $0{-}40\%$ centrality Pb+Pb collisions at the LHC \cite{Wilde:2012wc}. Both values are significantly larger than the critical temperature for chiral restoration and color deconfinement, $T_\mathrm{c}{\,\simeq\,}155-170$\,MeV \cite{Aoki:2006br,Bazavov:2011nk}, and hydrodynamic model studies reported in \cite{Adare:2008ab,Wilde:2012wc} show that the observations are consistent with much higher true fireball temperatures at a very early stage of the expansion of the collision fireball. The fact of $T_\mathrm{eff}$ being larger than $T_\mathrm{c}$ in itself, however, does not prove that the radiation was emitted from a quark-gluon plasma: It could, in principle, be due to radiating hadrons in the late stages of the collision where the true fireball temperature is already below $T_\mathrm{c}$ but strong radial flow boosts the emission spectrum to an effective temperature $T_\mathrm{eff}{\,>\,}T_\mathrm{c}$.
%
\begin{figure*}[hbt]
\begin{center}
    \includegraphics[width=0.47\linewidth]{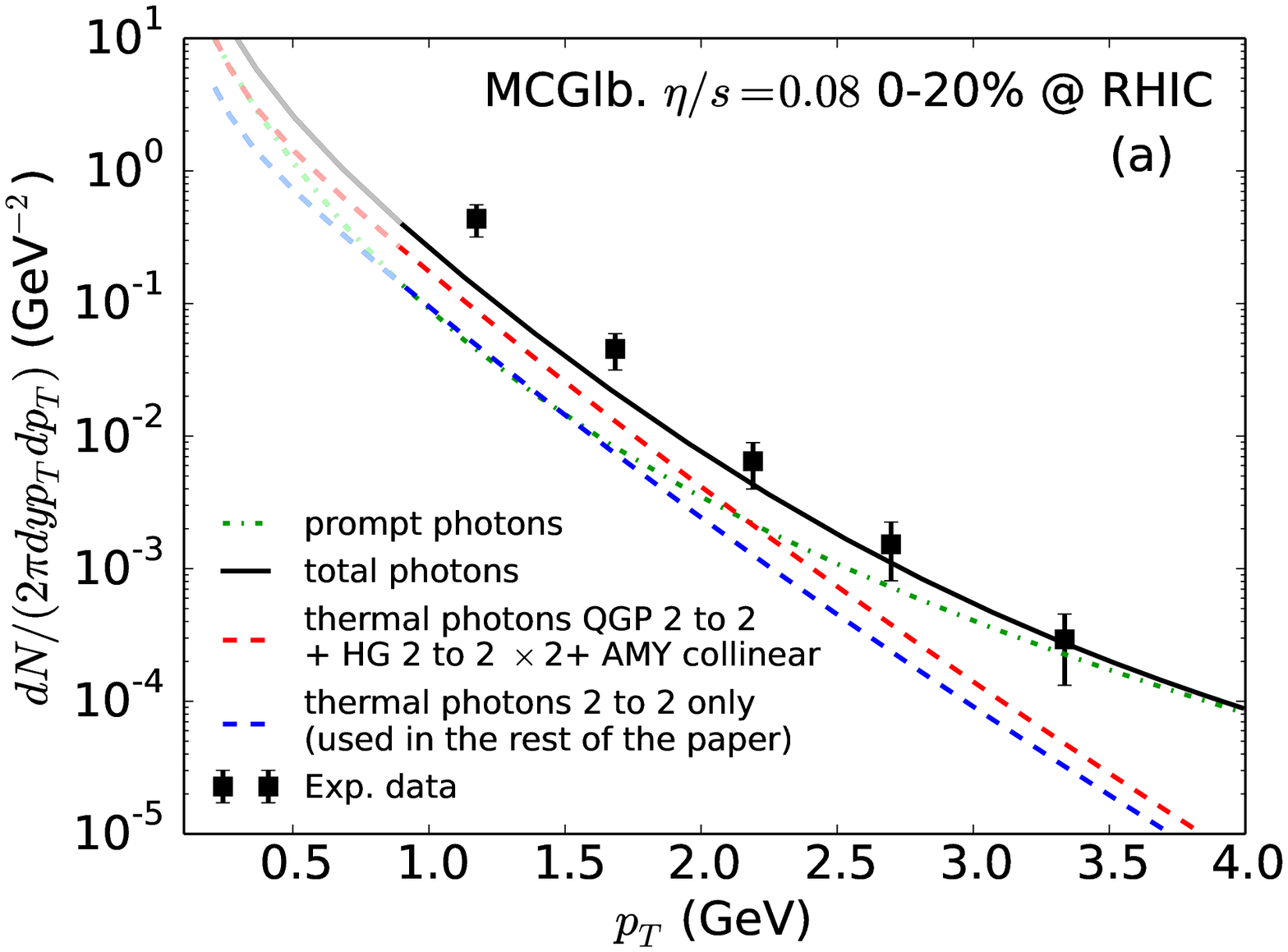}
    \includegraphics[width=0.47\linewidth]{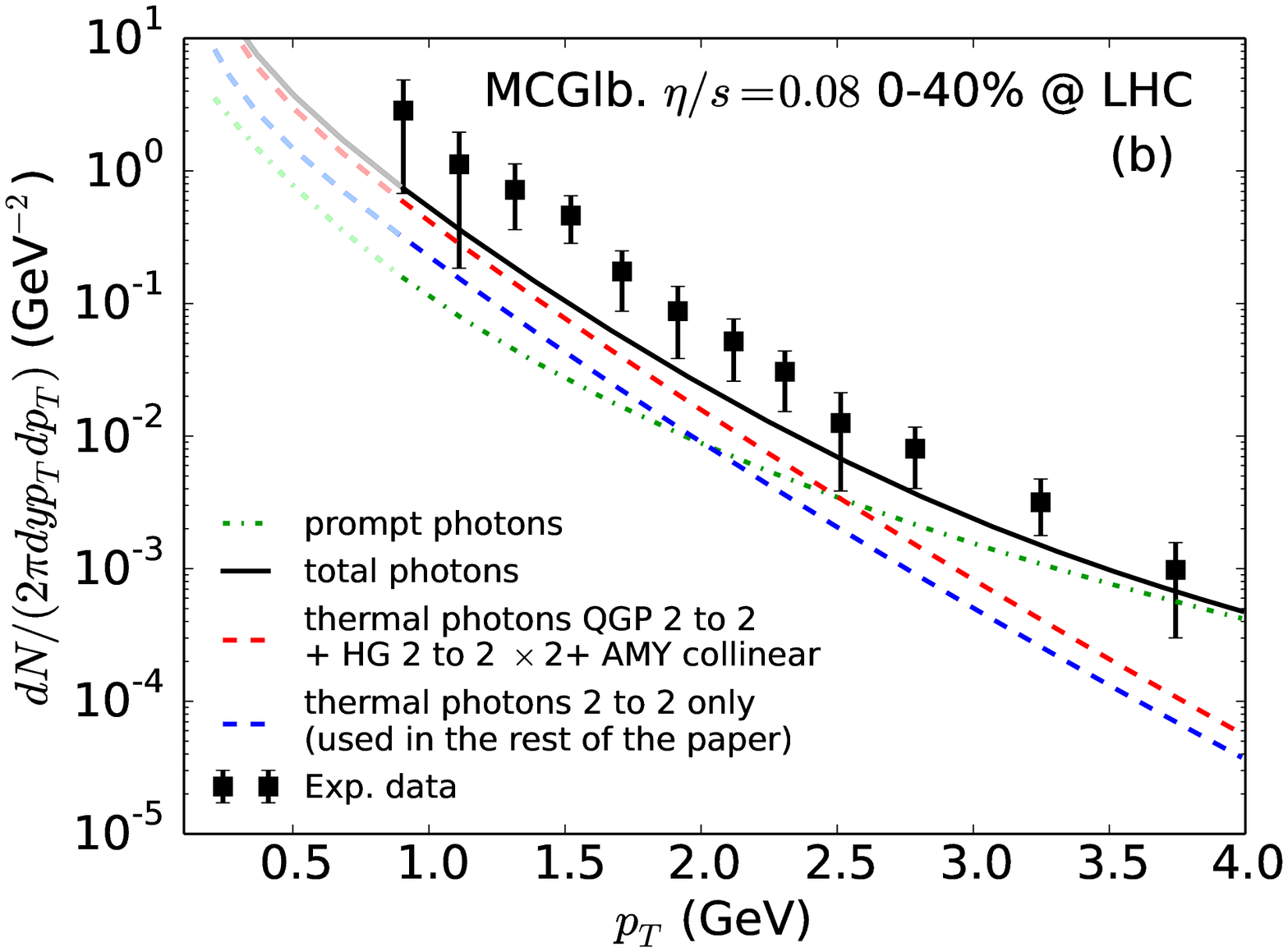}
    \caption{(Color online)
    Measured and calculated photon spectra in $0{-}20\%$ centrality Au+Au collisions at RHIC (a) 
    and $0{-}40\%$ centrality Pb+Pb collisions at the LHC (b). Photons from thermal sources and from 
    pQCD are shown separately, as well as their sum. The Au+Au collisions data at RHIC (a)
    are from the PHENIX Collaboration \cite{Adare:2008ab}, those for Pb+Pb collisions at the LHC (b) 
    from the ALICE Collaboration \cite{Wilde:2012wc}. See text for a detailed discussion. The shaded curves below 1 GeV are to remind of the uncertainties in extrapolating pQCD to low values of the photon transverse momentum \cite{pQCD}.
    \label{Pho_spectra}
    \vspace*{-5mm}
    }
\end{center}
\end{figure*}
%
Arguments for predominantly late emission of thermal photons, with momentum distributions that are strongly affected by collective flow, have been previously presented in \cite{vanHees:2011vb,Rapp:2011is}, based on a simple fireball evolution model with parameters constrained by hadronic observables. In this  work we use a realistic hydrodynamic simulation of the fireball evolution to explore the effects of hydrodynamic flow on the effective temperature (inverse slope) of the emitted thermal photon spectrum more quantitatively. Both our evolution and photon emission rates \cite{Shen:2013cca} incorporate viscous effects. We also study schematically the consequences of a hypothetical scenario where the fireball medium initially consists entirely of gluons (which do not radiate electromagnetically) and quark-antiquark creation (chemical equilibration) is delayed by several fm/$c$ \cite{Biro:1993qt,Gelis:2004ep}. How much will theoretical and experimental precision have to improve to allow to distinguish empirically between an initially ``dim'' gluon plasma and a QGP that reaches chemical equilibrium very quickly? In an attempt to start answering questions such as these, and to exploit the penetrating nature of the electromagnetic radiation, the  space-time history of photon emission is explored. We show that strategic cuts on the photon transverse momentum have the potential to make their thermometric nature even more explicit.

%
\begin{figure*}[htb]
\begin{center}
    \includegraphics[width=0.47\linewidth]{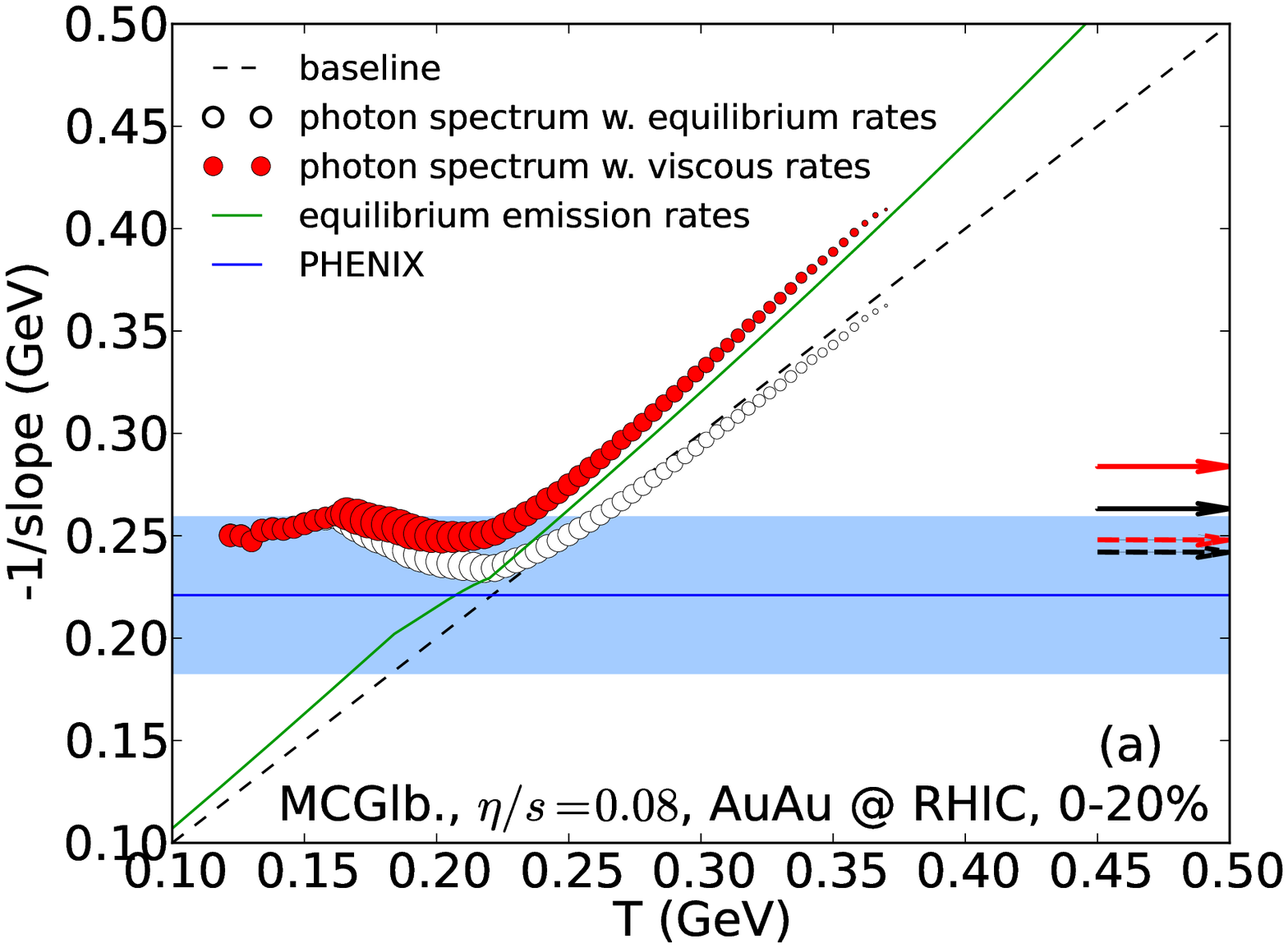}
    \includegraphics[width=0.47\linewidth]{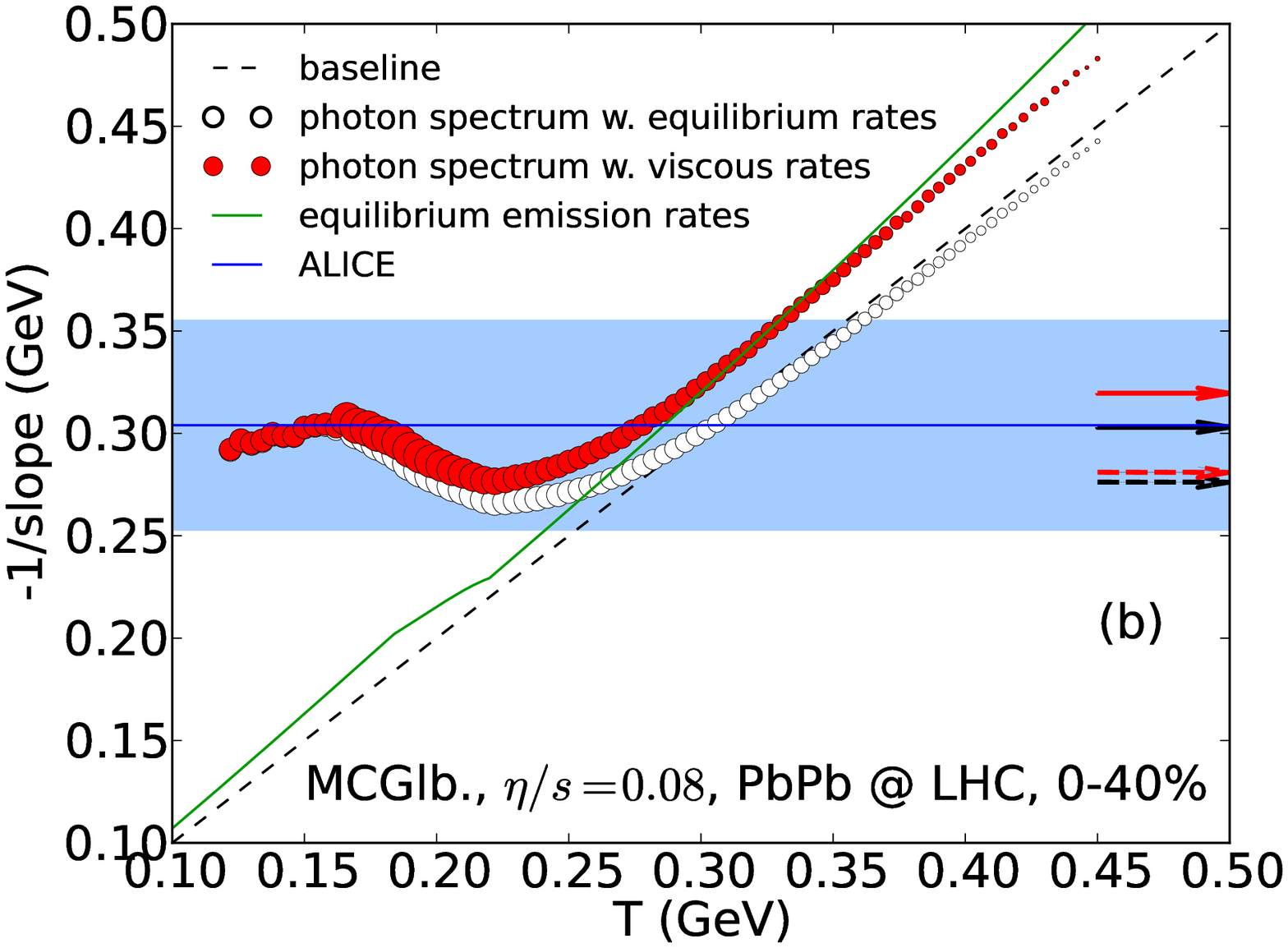}
    \caption{(Color online)
    The inverse photon slope parameter $T_\mathrm{eff}{\,=\,}{-}1/\mathrm{slope}$ as a function 
    of the local fluid cell temperature, from the equilibrium thermal emission rates (solid green
    lines) and from hydrodynamic simulations (open and filled circles), compared with the 
    experimental values (horizontal lines and error bands), for (a) Au+Au collisions at RHIC 
    and (b) Pb+Pb collisions at the LHC. The experimental values and error bands are from 
    the PHENIX Collaboration \cite{Adare:2008ab} in (a) and from the ALICE Collaboration 
    \cite{Wilde:2012wc} in (b). We note that the corresponding plot for Au+Au collisions at $20{-}40\%$ 
    centrality looks very similar to the case of $0{-}20\%$ centrality shown in panel (a), in agreement
    with what the PHENIX Collaboration \cite{Adare:2008ab} found. See text for a detailed discussion.  
    \label{F1}
    \vspace*{-5mm}
    }
\end{center}
\end{figure*}
%

The dynamical evolution of the radiating fireball is modeled with the boost-invariant hydrodynamic code {\tt VISH2{+}1} \cite{Song:2007fn}, using parameters extracted from previous phenomenologically successful studies of hadron production in 200\,$A$\,GeV Au+Au collisions at RHIC \cite{Shen:2010uy,Song:2010mg} and in 2.76\,$A$\,TeV Pb+Pb collisions at the LHC \cite{Shen:2011eg,Qiu:2011hf}. We here use ensemble-averaged Monte-Carlo Glauber (MCGlb) initial conditions which we propagate with $\eta/s{\,=\,}0.08$ \cite{Shen:2010uy,Song:2010mg,Shen:2011eg,Qiu:2011hf} and the lattice-based equation of state (EoS) s95p-PCE-v0 \cite{Huovinen:2009yb} which implements chemical freeze-out at $T_\mathrm{chem}{\,=\,}165$\,MeV. We start the hydrodynamic evolution at  $\tau_0{\,=\,}0.6$\,fm/$c$, corresponding to a peak initial temperature (energy density) in the fireball center of $T_0{\,=\,}452$\,MeV ($e_0{\,=\,}62$\,GeV/fm$^3$) at the LHC (Pb+Pb at $0{-}40\%$ centrality), and of $T_0{\,=\,}370$\,MeV ($e_0{\,=\,}35$\,GeV/fm$^3$) at RHIC (Au+Au at $0{-}20\%$ centrality). We end it on an isothermal hadronic freeze-out surface of temperature $T_\mathrm{dec}{\,=\,}120$\,MeV. 

Photons are emitted from the fireball using photon emission rates that are corrected \cite{Shen:2013cca} for deviations from local thermal equilibrium caused by the non-zero shear viscosity of the medium. We keep all terms linear in the viscous pressure tensor $\pi^{\mu\nu}$, both in the in- and outgoing distribution functions and in the self-energies of the particles exchanged in the radiative collision processes. At this point we include only $2{\,\to\,}2$ scattering processes; in the QGP our $2{\,\to\,}2$ rates are accurate to leading order of the strong coupling constant \cite{Shen:2013cca}. (A complete leading-order calculation including soft collinear gluon emission and its viscous corrections is under way.) We focus on photons with $p_T{\,<\,}4$\,GeV and ignore the contributions from hard pre-equilibrium processes which do not significantly affect the extraction of the inverse photon slope in this $p_T$-region \cite{Klasen:2013mga}. The hadronic phase (HG) is modeled as an interacting meson gas within the SU(3)${\,\times\,}$SU(3) massive Yang-Mills approach (see Refs.~\cite{Song:1993ae,Turbide:2003si,Turbide:2006zz} for details), with non-equilibrium chemical potentials to account for chemical decoupling at $T_\mathrm{chem}{\,=\,}165$\,MeV. Both approaches to computing the emission rates are expected to break down in the phase transition region. To avoid discontinuities, the QGP and HG emission rates are linearly interpolated in the temperature window 184\,MeV${\,<\,}T{\,<\,}220$\,MeV where our employed EoS \cite{Huovinen:2009yb} interpolates continuously between the lattice QCD data and the hadron resonance gas model in such a way that the smooth crossover character of the phase transition seen on the lattice is preserved.

As a prelude to the temperature studies to be reported in this work, it is useful to compare the results of our calculations with the photon spectra measurements performed at RHIC and at the LHC by the PHENIX and ALICE collaboration, respectively. The calculated spectra shown in Figure \ref{Pho_spectra} include the thermal rates corrected for shear viscosity effects integrated over the viscous hydrodynamical space-time evolution, and also the prompt photons resulting from the very early interactions of the partons distributed inside the nucleus. The photon rates used for these spectrum calculations differ slightly from what is used in the rest of this paper, due to the sensitivity of the photon spectra to details of the rates that are not relevant for the study of effective temperatures. For the QGP contribution shown in Fig. \ref{Pho_spectra}, the full leading order \emph{ideal}  rate for collinear emission is added to the viscous-corrected  $2\to 2$ rate. This is justified by the fact that the full leading order rate is known to have a similar energy dependence as the $2\to 2$ one (see e.g. \cite{Dion:2011pp}), but is about twice as large. This normalisation does not affect significantly effective temperature studies, but would lead to an artificial underestimate of the QGP photons that we want to avoid here. On the hadron gas side, we note that our rates do not include the contribution of baryons to the production of real photons.  As the baryonic contribution is roughly equal to that of the mesons at photon transverse momentum of $p_T{\,\approx\,}1$\,GeV/$c$ \cite{Turbide:2003si}, and as the baryonic rates are not yet amenable to a form which enables  viscous corrections, the photon yield from the hadronic medium is approximated here by multiplying the net mesonic contribution by a factor two. We emphasize that this normalisation of the hadron gas rate, along with the replacement of $2\to 2$ ideal QGP rate by the full leading order one, is only used here and not in what follows. The prompt photon calculation is performed at next-to-leading-order (NLO) in the strong coupling constant  \cite{Aurenche:2006vj,paquet}, and the nuclear parton  distribution functions are corrected for isospin and shadowing effects \cite{Eskola:2009uj}. The calculation is extrapolated to low transverse momentum using a fit of the form $A/(1+p_T/p_0)^{n}$, a form that we checked describes very well the available low $p_T$ photon data from proton-proton collisions \cite{Adare:2008ab,pQCD} For RHIC and LHC conditions, the agreement between the calculated results and measured data shown here is not untypical of that obtained with other contemporary hydrodynamic simulations \cite{Turbide:2007mi,Qin:2009bk,Chatterjee:2013naa}. We note that the data allow room for additional photon sources in addition to the ones considered here. Among the possible candidates are photons from jet-plasma interactions \cite{jet-pho,Qin:2009bk}; one should also keep in mind the role played by fluctuating initial states \cite{Chatterjee:2011dw,Dion:2011pp}.

%
\begin{figure*}[htb]
\begin{center}
    \includegraphics[width=0.47\linewidth]{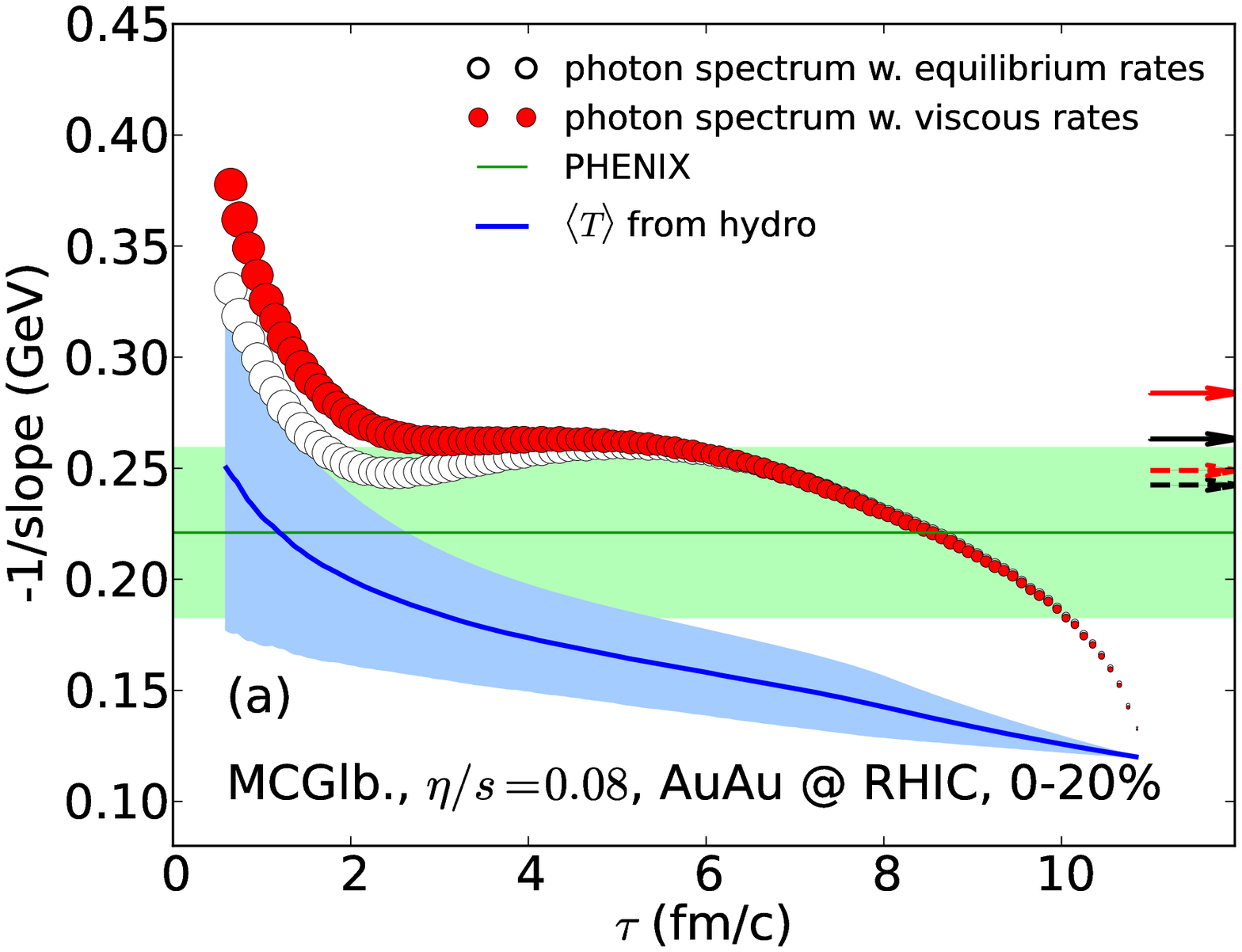}
    \includegraphics[width=0.47\linewidth]{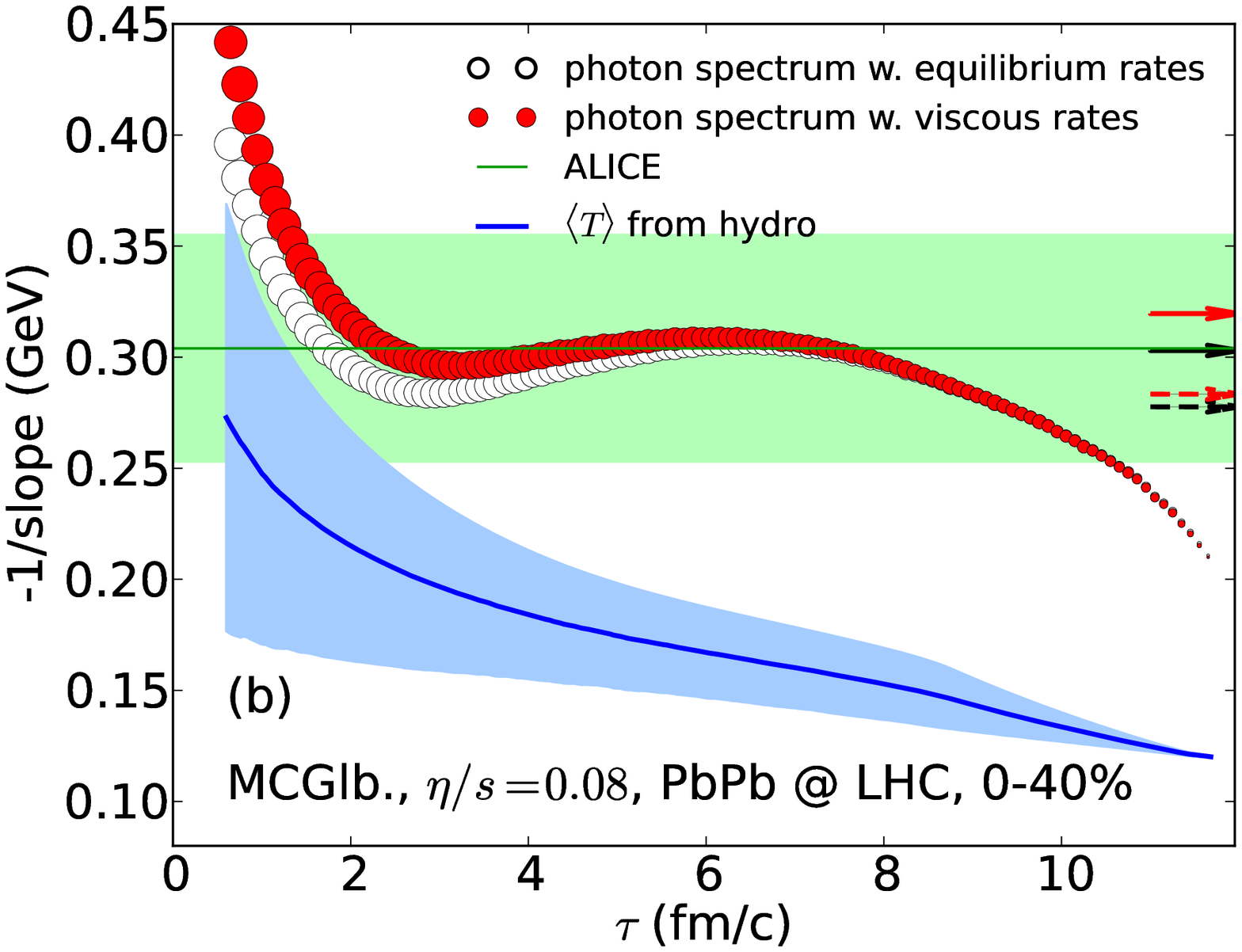}
    \caption{(Color online)
    The inverse photon slope parameter $T_\mathrm{eff}{\,=\,}{-}1/\mathrm{slope}$ as a 
    function of emission time from hydrodynamic simulations, compared with the experimental
    (time-integrated) values (horizontal lines and error bands), for (a) Au+Au collisions at 
    RHIC and (b) Pb+Pb collisions at the LHC. The blue solid lines and surrounding shaded 
    areas show for comparison the time evolution of the average fireball temperature and its 
    standard deviation. See text for further discussion.  
    \label{F2}
    \vspace*{-5mm}
    }
\end{center}
\end{figure*}
%

The equilibrium emission rates as well as the non-equilibrium photon spectra emitted during the hydrodynamic evolution are approximately exponential in $p_T$ between 1 and 4\,GeV, and we can characterize them by their inverse logarithmic slopes $T_\mathrm{eff}$ just as the experiments did for the measured photon spectra. The green lines in Fig.~\ref{F1} show $T_\mathrm{eff}$ as a function of the true temperature $T$ for the equilibrium photon emission rates. One sees that, due to phase-space factors associated with the radiation process, the effective temperature of the emission rate is somewhat larger than the true temperature: At high $T$, the QGP photon emission rate goes roughly as $\exp(-E_\gamma/T)\log(E_\gamma/T)$ \cite{Kapusta:1991qp}, and the logarithmic factor is responsible for the somewhat harder emission spectrum. The double kink in the green line at $T{\,=\,}184$ and 220\,MeV reflects the interpolation between the QGP and HG rates. The effect of that interpolation on the slope of the spectrum is weak, one mainly interpolates between rates with different normalization. 

The circles in Fig.~\ref{F1} show the effective temperatures of photons emitted with equilibrium rates (open black circles) and with viscously corrected rates (filled red circles) from cells of a given temperature within the hydrodynamically evolving viscous medium. The area of the circles is proportional to the total photon yield emitted from all cells at that temperature. One sees here and also in Fig.~\ref{F2} below that viscous corrections to the photon emission rates are large at early times (high $T$), due to the initially very large longitudinal expansion rate, but become negligible at later times (lower $T$). Viscous effects on the emission rates harden the photon spectrum (i.e. they increase $T_\mathrm{eff}$) but do not affect the photon yields. The hydrodynamic photon spectra using ideal rates (open black circles) have lower effective temperatures than the local emission rates themselves (green solid lines): This is due to the integration of the Boltzmann factor $e^{-E_\gamma/T}{\,=\,}e^{-p_T\cosh(y{-}\eta)/T}$  over space-time rapidity $\eta$ which, for fixed $T$, sums over contributions\footnote{Recall that we assume a boost-invariant 
               (i.e. $\eta$-independent) distribution of thermal sources.}
with different effective temperatures $T_\mathrm{eff}(\eta){\,=\,}\frac{T}{\cosh(y{-}\eta)}{\,<\,}T$ (we here consider photons at midrapidity, $y{\,=\,}0$). Surprisingly, this rapidity-smearing effect leads, for ideal emission rates, to photon spectra whose inverse slope reflects at early times almost exactly the temperature of the emitting fluid cells in their rest frame. Including viscous corrections in the emission rates increases the effective photon temperatures by about 10\% at early times.

As the system cools, Fig.~\ref{F1} shows that the effective photon temperature begins to deviate upward from the true temperature. Below $T{\,\sim\,}220$\,MeV the effective photon temperature actually begins to increase again while the true temperature continues to decrease. This is caused by the strengthening radial flow; below $T{\,\sim\,}220$\,MeV, the radial boost effect on $T_\mathrm{eff}$ overcompensates for the fireball cooling. Once the system reaches chemical freeze-out at $T_\mathrm{chem}{\,=\,}165$\,MeV, the character of the equation of state changes, leading to faster cooling \cite{Hirano:2002ds} without developing additional radial flow at a sufficient rate to keep compensating for the drop in effective temperature due to this cooling. The faster expansion below $T_\mathrm{chem}$ is also seen in the solid blue lines in Fig.~\ref{F2}, and it is reflected in the shrinking size of the circles (integrated photon yields) in Fig.~\ref{F1} below $T_\mathrm{chem}$, reflecting the smaller space-time volumes occupied by cells with temperatures $T{\,<\,}T_\mathrm{chem}$. 
 
%
\begin{figure*}[hbt!]
\begin{center}
    \includegraphics[width=0.45\linewidth]{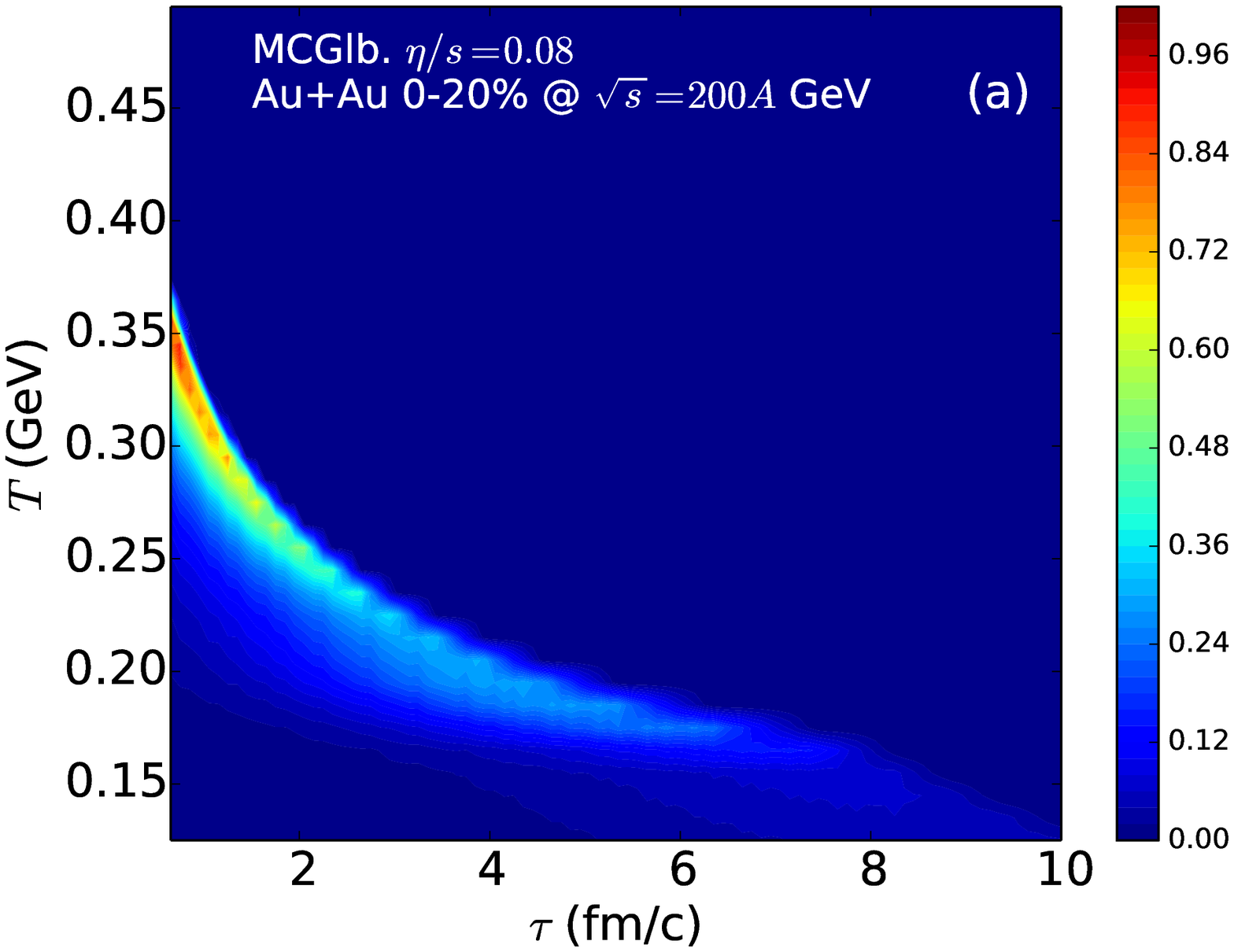}
    \includegraphics[width=0.45\linewidth]{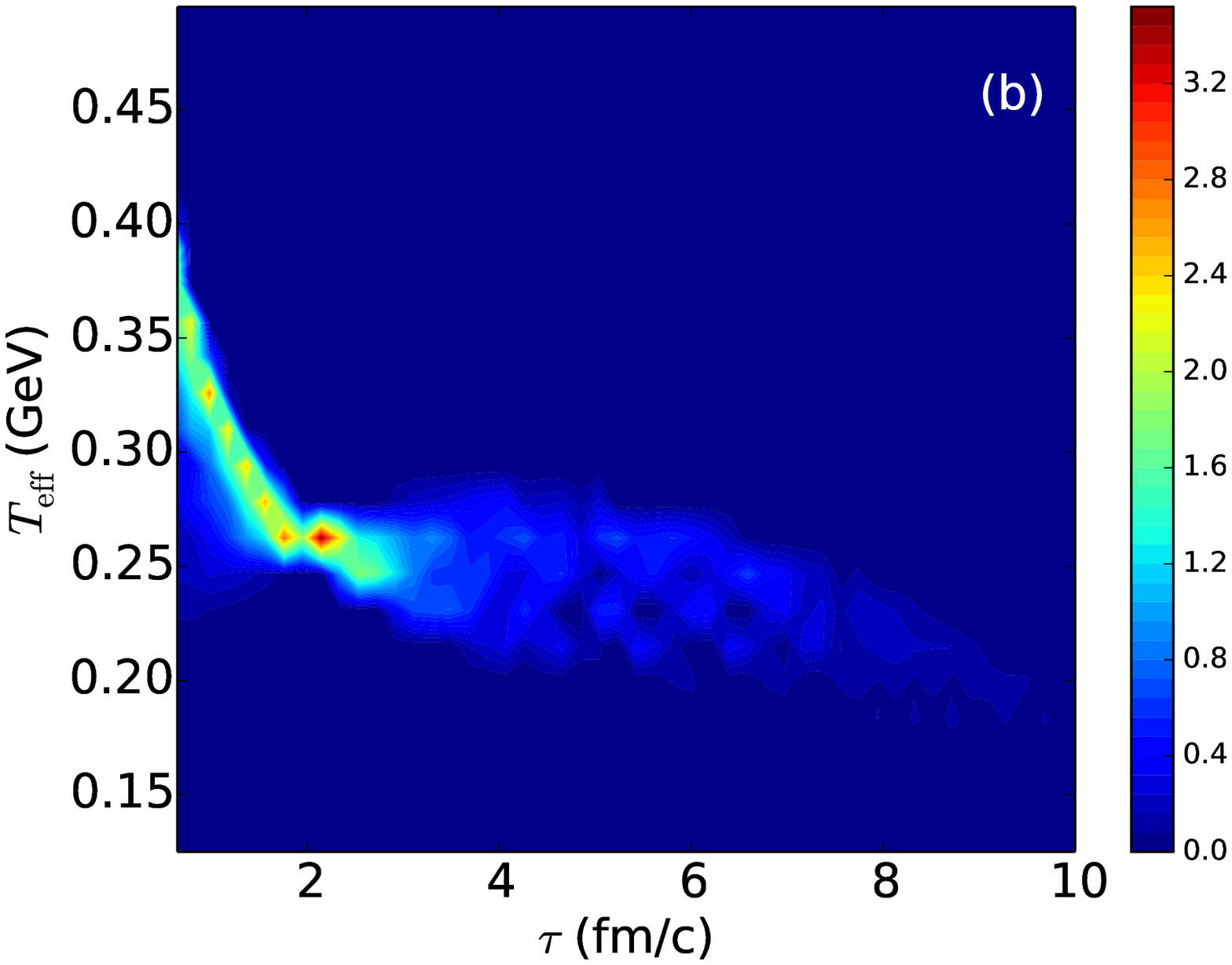}
    \includegraphics[width=0.45\linewidth]{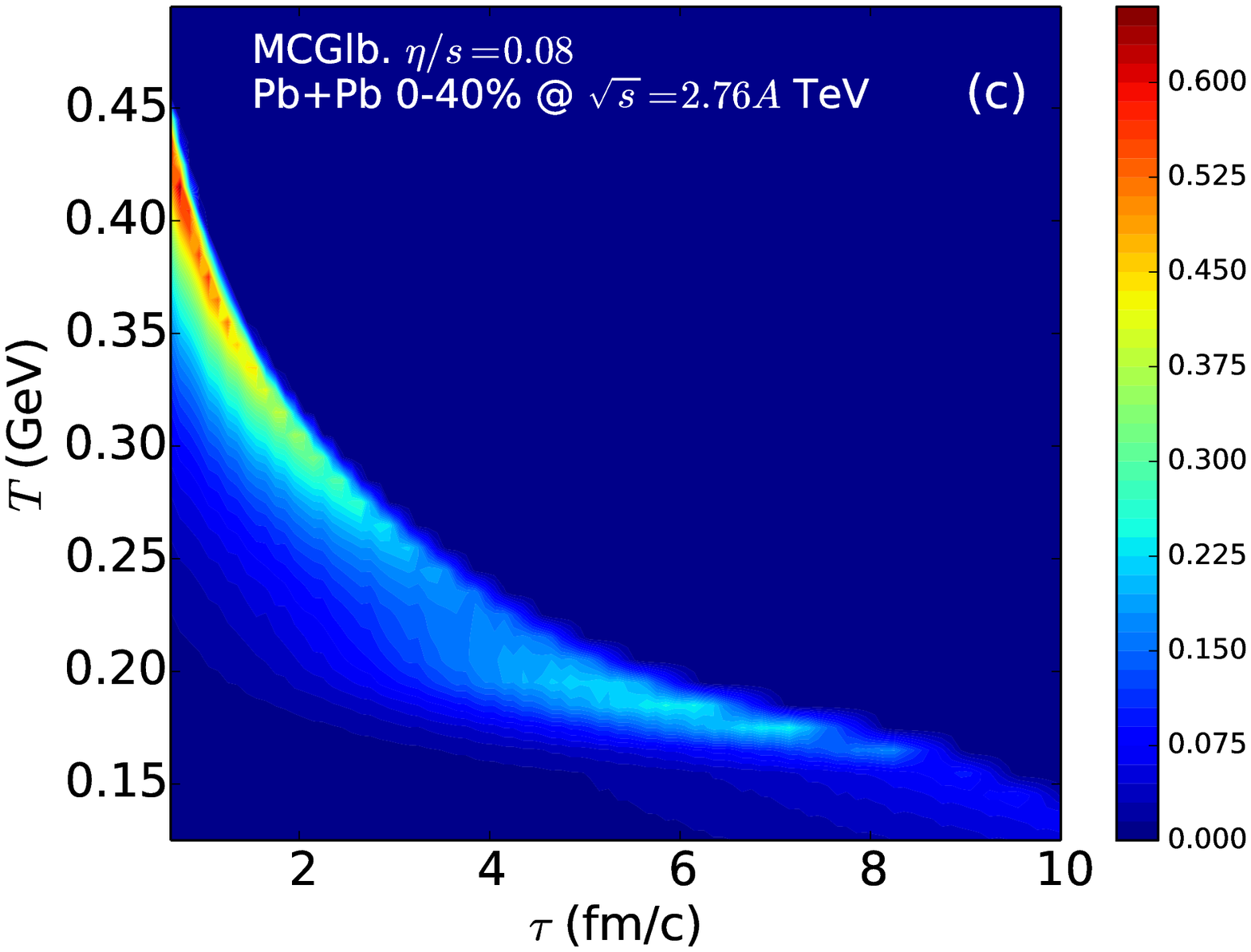}
    \includegraphics[width=0.45\linewidth]{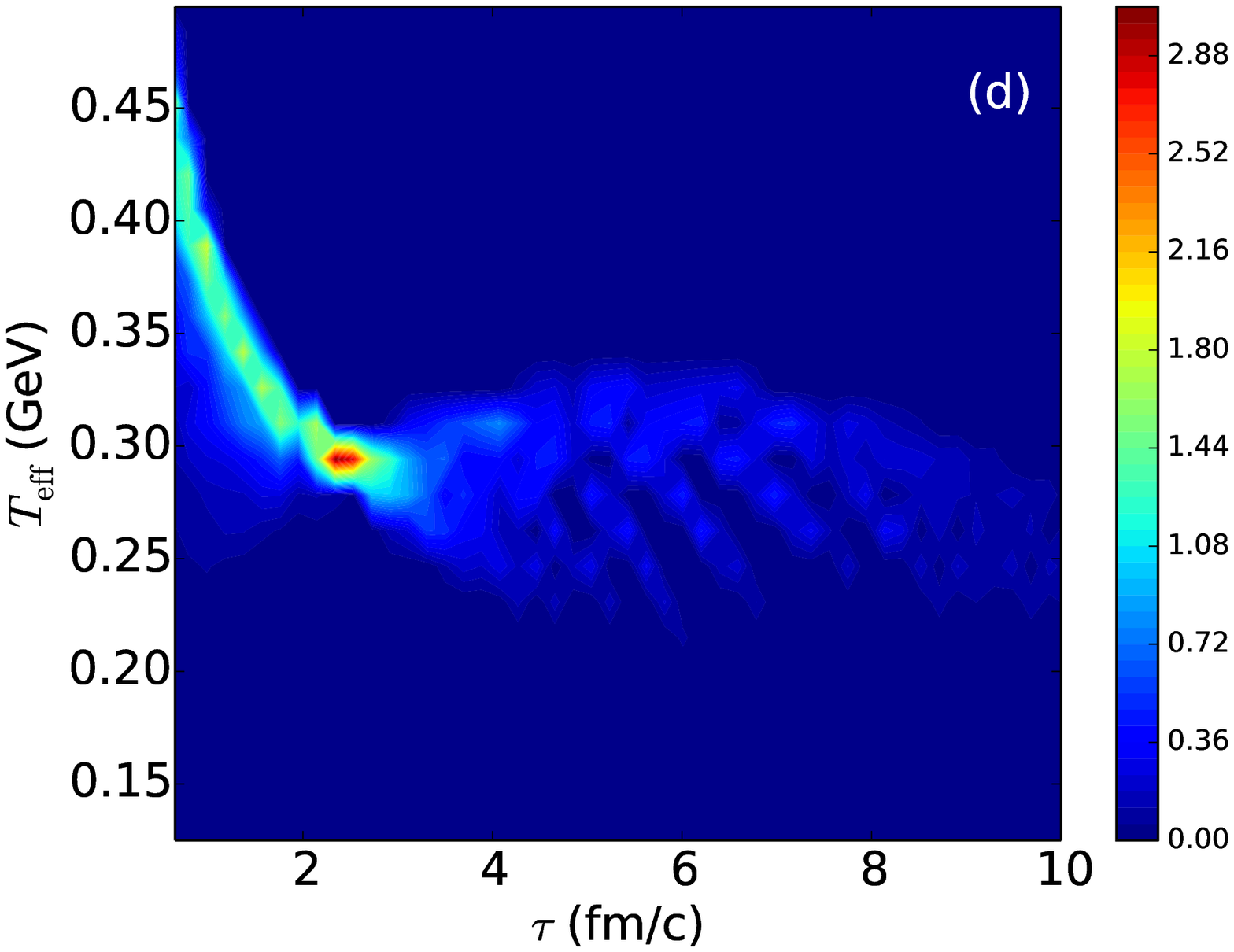}
    \caption{(Color online)
Contour plots of the normalized differential photon yield $\frac{dN^\gamma(T,\tau)/(dy\,dT\,d\tau)}{dN^\gamma/dy}$ (panels (a) and (c)) and $\frac{dN^\gamma(T_\mathrm{eff},\tau)/(dy\,dT_\mathrm{eff}\,d\tau)}{dN^\gamma/dy}$ (panels (b) and (d)) for Au+Au collisions at RHIC at 0-20\% centrality (panels (a) and (b)) and for Pb+Pb collisions at the LHC at 0-40\% centrality (panels (c) and (d)). The color bars translate the colors into absolute values (in $c/$(GeV\,fm)) for the quantities plotted. See text for discussion. 
    \label{F3}
    \vspace*{-5mm}
    }
\end{center}
\end{figure*}
%

Fig.~\ref{F2} shows the effective slopes of photons emitted at different times from the expanding fireball, again compared with the time-integrated experimental values (horizontal bands). (A similar graph, based on a parametrized fireball evolution model with thermal equilibrium rates and a first-order phase transition, can be found in Fig.\,7 in Ref.~\cite{vanHees:2011vb}). As before, the open black circles use equilibrium emission rates while the filled red circles account for viscous corrections to the photon emission rates. (The hydrodynamic expansion is viscous in both cases.) For comparison, the blue lines show the evolution of the average fireball temperature (averaged over all cells with $T{\,>\,}120$\,MeV at time $\tau$), with shaded regions indicating its standard deviation. After about 2\,fm/$c$, the effective photon temperature begins to get significantly blue-shifted  by radial flow. This radial boost is clearly stronger at the LHC than at RHIC. Radial flow effects decrease again at very late times when only a small region near the fireball center survives where the radial flow goes to zero. The difference between open and filled circles shows that viscous effects on the photon emission rates are concentrated at early times.

While Fig.~\ref{F2} demonstrates that  the early photons are associated with a high  yield (as is commonly understood), Fig.~\ref{F1} shows that most photons are emitted from a relatively narrow temperature band between 165 and 220 MeV. Relatively few of the photons thus come from the hot core of the fireball; a much larger fraction comes from the cooler periphery and is emitted with temperatures close to the quark-hadron transition. Averaged over time, these photons from the transition region are strongly affected by radial flow, resulting in inverse slopes (``effective temperatures'') that are much larger than their true emission temperatures.  These findings can even be put on a firmer quantitative basis, by considering the following: At each value of proper time, $\tau$, photons are emitted with a distribution of thermodynamic temperatures. This distribution is shown in Figs.~\ref{F3}a (for Au+Au at RHIC) and \ref{F3}c (for Pb+Pb at the LHC), where the color-coding of the contour plots reflects the differential photon yield (normalized to the total yield $dN^\gamma/dy$) per time and temperature (in $c/$(GeV\,fm)) in the $T{-}\tau$ plane. The corresponding distribution of flow-blue-shifted effective temperatures $T_\mathrm{eff}$ (inverse slopes) is shown in Figs.~\ref{F3}b (for RHIC) and \ref{F3}d (for the LHC). Comparing the left and right panels one observes, after a proper time $\tau{\,\sim\,}2$\,fm/$c$, a clear shift to higher effective temperatures, owing to the development of radial hydrodynamic flow.  Furthermore, the dependence of the effective temperature on the flow velocity (which depends on radial position)  leads to an additional broadening of the distribution of $T_{\rm eff}$ at any given time. 

%
\begin{figure*}[thb!]
\begin{center}
    \includegraphics[width=0.47\linewidth]{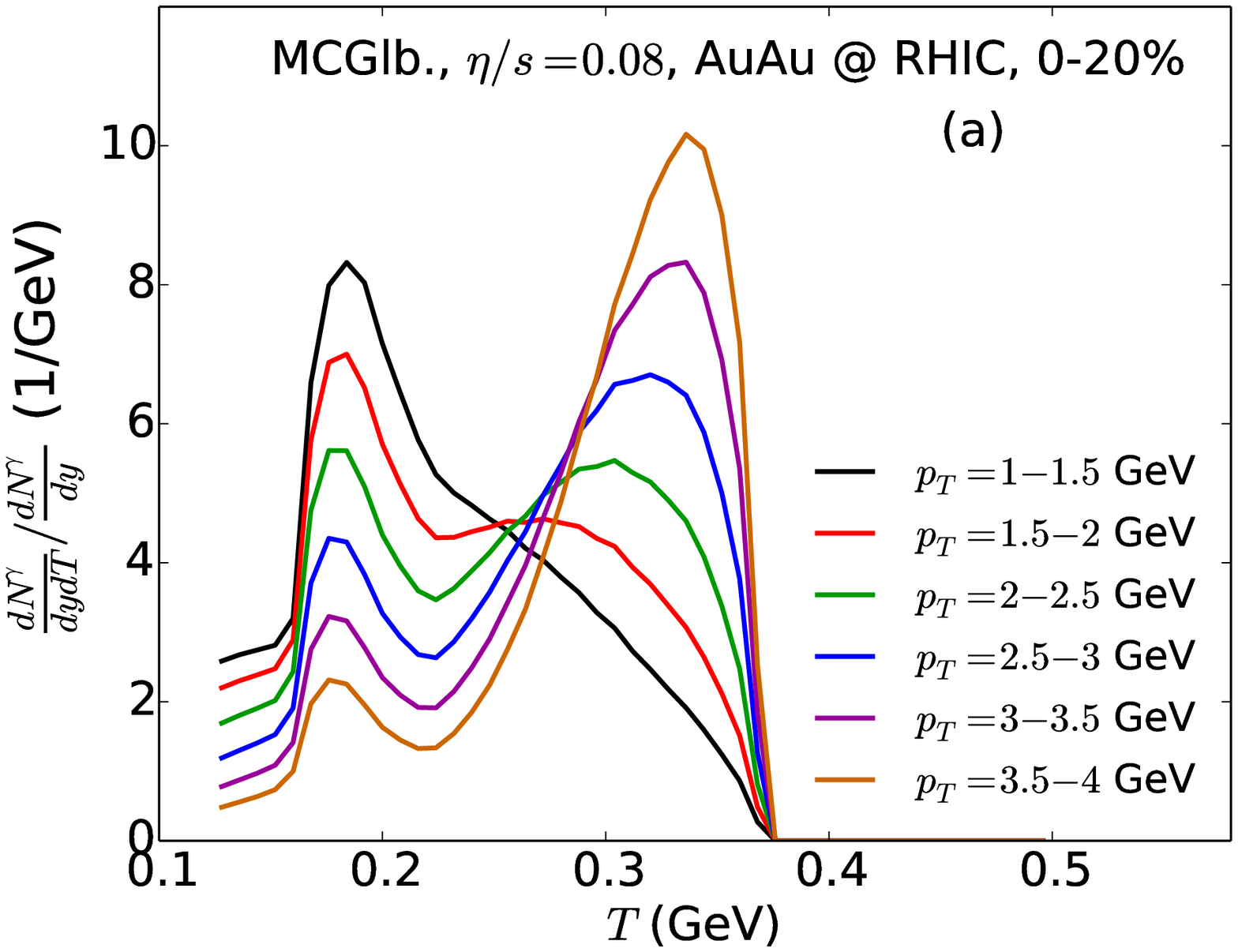}
    \includegraphics[width=0.47\linewidth]{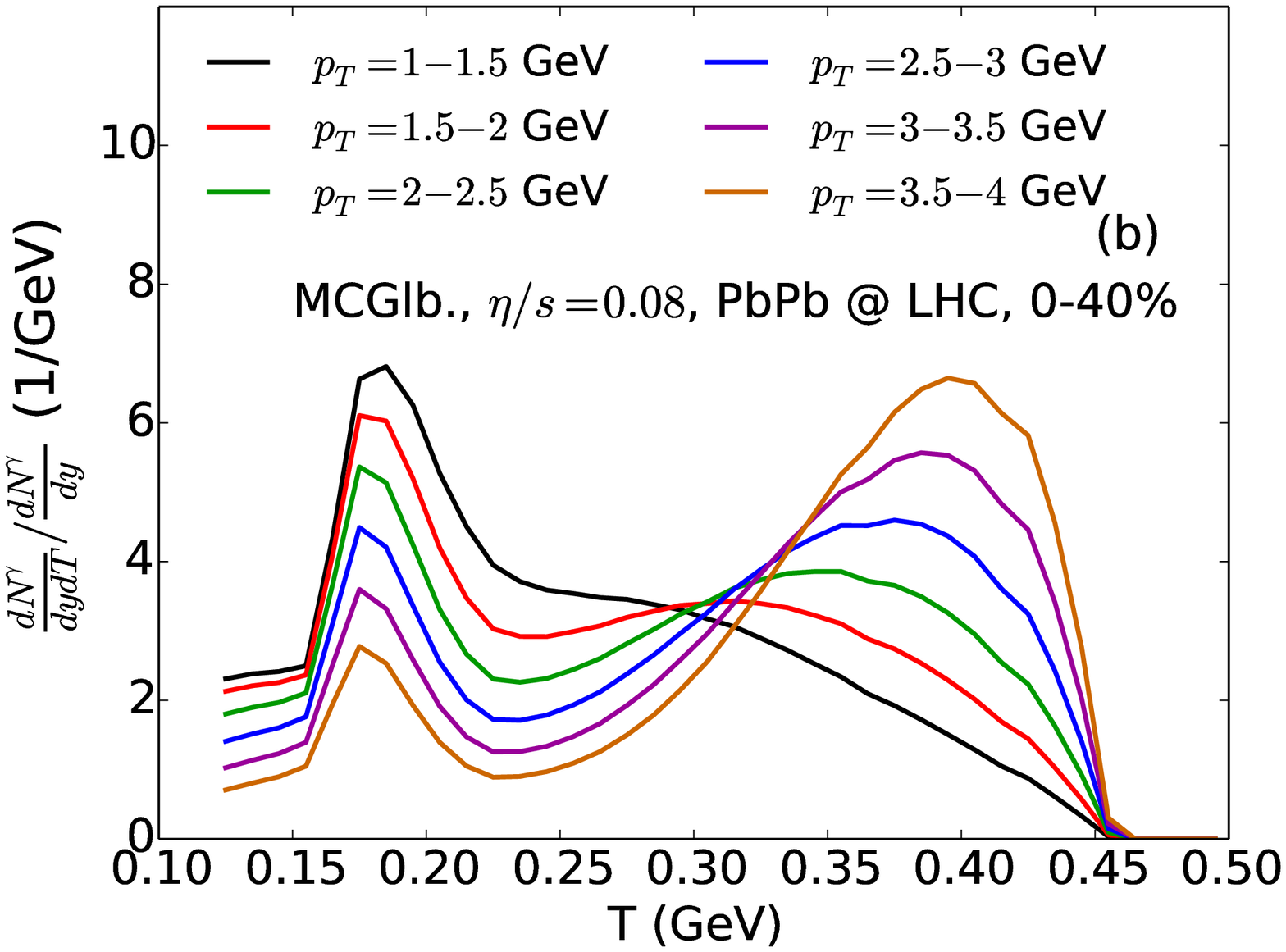}
    \caption{(Color online)
    The differential photon yield, as a function of the temperature $T$, for different windows in photon transverse momentum,  for Au+Au collisions at 
    RHIC (a) and  Pb+Pb collisions at the LHC (b).  
    \label{F4}
    \vspace*{-5mm}
    }
\end{center}
\end{figure*}
%

In order to further quantify the connection between the photon spectrum and the emission temperature, a model calculation allows to dissect the photon contribution in terms of  transverse momentum. Figure \ref{F4} shows the relative photon yield in different transverse momentum regions, as a function of the temperature at which those photons were radiated. The photon yield is obtained by integrating the flow-boosted photon emission rate over the space-time volume. The rate is large at high temperatures, but the corresponding space-time volume is small. As the system cools and the rate drops, the decrease in the rate is (partially) offset by the increasing fireball volume, and the shift to lower photon energies resulting from the cooling is counteracted by the increasing radial flow. The combination of these effects can create a bi-modal distribution of the thermodynamic temperatures that contribute to photon production in a given $p_T$ window. The relative size of the two peaks corresponding to emission from very hot cells with little flow and rather cool cells with strong radial flow depends on the photon momentum, as is shown in Fig.~\ref{F4}. The curves in Fig.~\ref{F4} can be viewed as thermal photonsÕ probability distribution function, with the local temperature as the independent variable. At RHIC, one observes that for $p_T{\,=\,}1{-}1.5$\,GeV, photons are mostly coming from the transition region $T{\,=\,}150{-}220$\,MeV. For harder thermal photons with $p_T{\,=\,}3.5{-}4$\,GeV, on the other hand, the emission probability is peaked at temperatures above 300\,MeV, followed by a second wave of emission from a much larger space-time volume at temperatures around $T{\,=\,}150{-}220$\,GeV, blue-shifted by large radial flow. Interestingly, this bimodal structure does not require a first order phase transition but, as can be seen here, is also observed for a smooth cross-over. The right panel in Fig.~\ref{F4} shows the situation for conditions specific to the LHC. This figure illustrates well the power and advantages related to the use of a quantitative space-time modeling of the nuclear collisions: the momentum cuts shown here can be used to guide experimental analyses and simulations that seek to extract precise temperature information from high-energy nuclear collisions\footnote{Note however, that the high-momentum bins will receive pQCD contributions.}.

\begin{figure*}[bht!]
\begin{center}
   \includegraphics[width=0.8\linewidth]{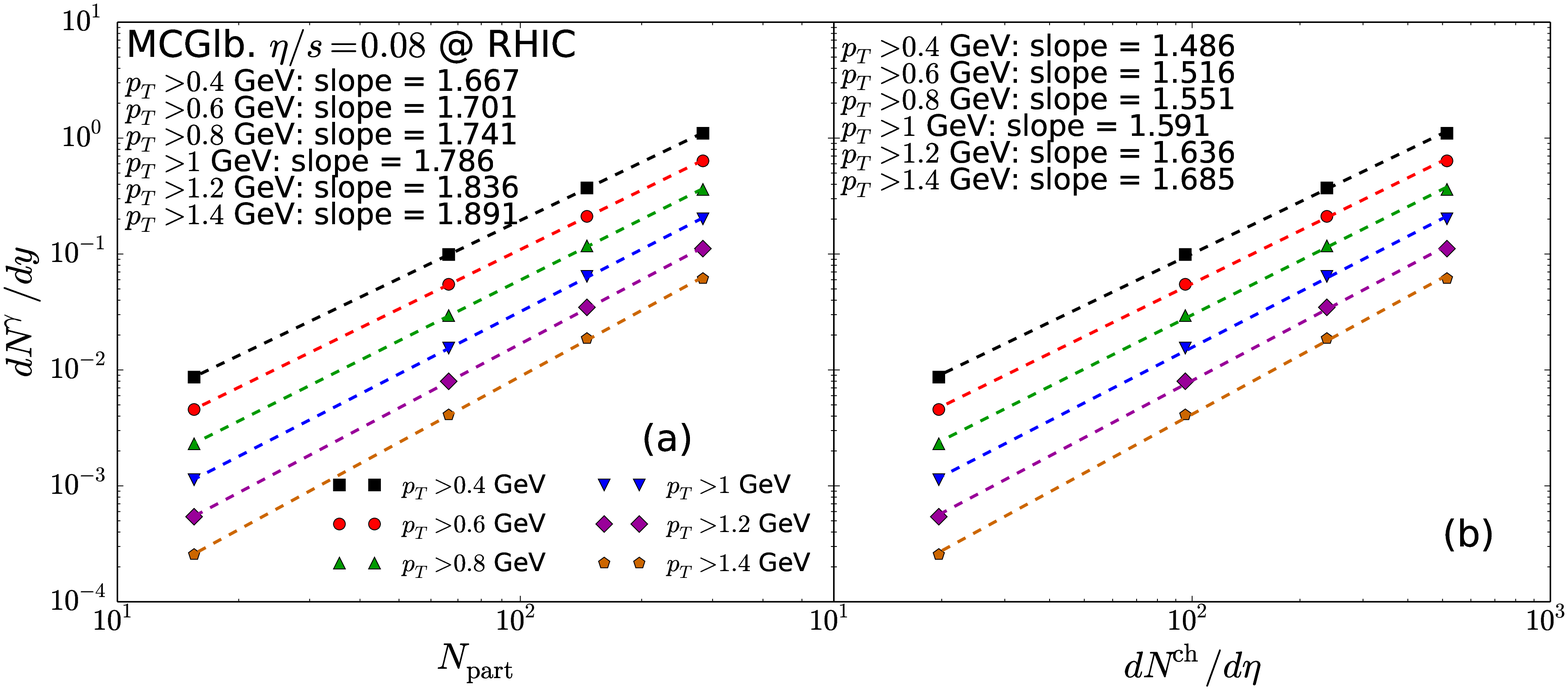}
    \caption{(Color online)
    The centrality dependence of the photon yield, for Au+Au collisions at RHIC. The centrality may be expressed in terms of (a) $N_{\rm part}$ or of (b) $d N^{\rm ch}/ d\eta$. 
    \label{F5}
    \vspace*{-5mm}
    }
\end{center}
\end{figure*}

Our hydrodynamical treatment also allows us to study the centrality dependence of the thermal photon yields (see also \cite{Linnyk:2013wma}), recently reported by the PHENIX Collaboration \cite{Bannier_HP2013}. That thermal photons should exhibit a stronger centrality dependence than hadrons and photons from hadronic decays is a very old idea that goes back to the beginning of the field of relativistic heavy-ion collisions \cite {Feinberg:1976ua,Hwa:1985xg}. Our studies show that a similar difference in centrality dependence is observed for thermal electromagnetic radiation from the QGP and HG phases. Figure~\ref{F5} illustrates this for RHIC conditions. We integrate the thermal photon spectra from a variable lower $p_T$ cutoff to a fixed upper limit of $p_T{\,=\,}4$\,GeV and plot the result in Fig.~\ref{F5} as function of centrality, measured either through $N_\mathrm{part}$ in panel (a) or through the charged multiplicity $dN_\mathrm{ch}/d\eta$ in panel (b). The latter plot permits a direct comparison with experimental measurements. The four points in each curve, from right to left, represent 0-20\%, 20-40\%, 40-60\%, and 60-95\% centralities. The dashed lines are  power-law fits to the points. One observes a thermal photon yield that grows like a power of $N_{\rm part}$ and as a (different) power of the multiplicity, with exponents stronger than linear. The powers depend on the lower $p_T$ cutoff with which the yield is evaluated. For the region $p_T{\,>\,}0.6$ GeV, the slope in the logarithmic plot of Fig.~\ref{F5}a is 1.7, slightly above the experimentally measured value of $1.48 \pm 0.08 \pm 0.04$ \cite{Bannier_HP2013}.

In order to further explore the possible implications of this difference in slopes between theory and data, we also computed, for the case shown in Fig.~\ref{F5}, the centrality dependences of the total QGP and HG photon yields above $p_T{\,=\,}0.4$\,GeV. The QGP photon yield is defined as all photons from cells with $T{\,>\,}220$\,MeV plus the QGP fraction of photons from cells with temperatures between 184 and 220\,MeV where we linearly interpolate between the QGP and HG emission rates. The HG photon yield is the complement of the total photon yield with respect to the QGP photon yield. We find the HG photon yield above $p_T{\,=\,}0.4$\,GeV to scale as function of $N_{\rm part}$ with power 1.46 and as function of $dN_\mathrm{ch}/d\eta$ with power 1.23; the corresponding scaling powers for the QGP photons are larger, 2.05 and 1.83, respectively. QGP photons thus have a significantly stronger centrality dependence than HG photons. The experimentally measured centrality dependence of all thermal photons is closer to the power predicted in our calculations for HG photons than for QGP photons. Together with our observation (not shown here) that our calculations significantly under-predict the measured total thermal photon yields at all centralities, this may indicate that our hydrodynamic calculations underestimate the photon production rate in the HG phase and/or near the quark-hadron phase transition. This observation invites further scrutiny in terms of its sensitivity to variations in the initial conditions and the transport coefficients the expanding hydrodynamic fluid.  
%
%

We also investigated the centrality dependence of the inverse slope of the thermal photon spectra in Au+Au collisions at RHIC (see Table~\ref{T2}). 
%
\begin{table}[hbt]
\begin{center}
\begin{tabular}{|c|c|c|}
\hline \hline
Centrality & PHENIX preliminary  & $T_\mathrm{eff}$ \\ 
                & results (MeV)            &  (MeV) \\\hline
  0-20\% & $237{\,\pm\,}25{\,\pm\,}29$ &  267\\ \hline
20-40\% & $260{\,\pm\,}33{\,\pm\,}31$ &  259\\ \hline
40-60\% & $228{\,\pm\,}28{\,\pm\,}27$ &  246\\ \hline
60-92\% & $254{\,\pm\,}53{\,\pm\,}25$ &  225\\ \hline \hline
\end{tabular}
\end{center}
\caption{Preliminary results for the inverse slope parameters extracted from thermal photon spectra for 200\,$A$\,GeV Au+Au collisions obtained by the PHENIX Collaboration \cite{Bannier_HP2013}, compared with those from the hydrodynamic model, for different collision centralities. To facilitate comparison with the experimental data, the theoretical spectra where fitted to exponentials in $p_T$ in the same interval as used in \cite{Bannier_HP2013}, $0.6{\,<\,}p_T{\,<\,}2.0$\,GeV.
\label{T2}
}
\end{table}
%
For the hydrodynamic runs on which Fig.~\ref{F5} is based, our results show a very weak centrality dependence, with $T_\mathrm{eff}$ being slightly smaller in peripheral than central collisions.

Returning to Figs.~\ref{F1} and \ref{F2}, we see that the large measured values for the inverse photon slope reflect, on average, true emission temperatures that lie {\em well below} the observed effective temperature. This raises an interesting question: Could it be that in the experiments we don't see any photons {\em at all} from temperatures well above $T_\mathrm{c}$, and that all measured photons stem from regions close to $T_\mathrm{c}$ and below, blue-shifted by radial flow to effective temperature values above $T_\mathrm{c}$? To get an idea what the answer to this question might be we performed a schematic study where in Fig.~\ref{F1} we turned off by hand all contributions to the photon spectrum from cells with true temperatures above 220\,MeV at RHIC and above 250\,MeV at the LHC (corresponding to about 1/3 of the total photon yield in both cases), and in Fig.~\ref{F2} all contributions from $\tau{\,<\,}2$\,fm/$c$ (corresponding to 26\% and 28.5\% of the total photon yield for RHIC and LHC collisions, respectively, see Table~\ref{T1}).\footnote{This implements, in a very rough way, the idea
               that the initial fireball state might be purely gluonic, and that chemical equilibration of 
               quarks can be characterized by a time constant taken to be about 2\,fm/$c$. It ignores, 
               however, that an initial suppression of quarks must be compensated by an increase in the 
               gluon temperature \cite{Biro:1993qt,Gelis:2004ep}, in order to maintain the same total 
               entropy and final multiplicity. As quarks are being produced from gluons, these quarks
               thus radiate more strongly than in chemical equilibrium, leading to a cancellation that 
               leaves the total photon spectrum almost unchanged \cite{Gelis:2004ep}. Our simplified 
               treatment ignores this increase in temperature and thus overestimates the effect of 
               early-time quark suppression on the photon spectrum. In  this sense, our conclusion from 
               this study is conservative.}
We show as arrows pointing to the right vertical axes in Figs.~\ref{F1} and \ref{F2} the inverse slopes of the final space-time integrated hydrodynamic photon spectra: Solid black and red lines correspond to calculations assuming full chemical equilibrium from the beginning and using thermal equilibrium and viscously corrected photon emission rates, respectively. The dashed black and red arrows show the same for calculations with delayed chemical equilibration, as described above. 
%
\begin{table}[t]
\begin{center}
\begin{tabular}{|c|c|c|}
\hline
range of photon  & \multicolumn{2}{c|}{fraction of total photon yield} \\\cline{2-3}
 emission & AuAu@RHIC & PbPb@LHC \\
 & 0-20\% centr. & 0-40\% centr. \\ \hline 
$T = 120$-$165$\,MeV & 17\% & 15\% \\
$T = 165$-$250$\,MeV & 62\% & 53\% \\
$T > 250$\,MeV & 21\% & 32\% \\ \hline \hline
$\tau = 0.6-2.0$\,fm/c & 28.5\% & 26\% \\
$\tau > 2.0$\,fm/c & 71.5\% & 74\% \\ \hline
\end{tabular}
\end{center}
\caption{Fractions of the total photon yield emitted from the expanding viscous hydrodynamic fireball from various space-time regions as indicated, for the two classes of collisions considered in this work.}
\label{T1}
\end{table}
%
The (overestimated) effects of our schematic handling of delayed chemical equilibration on the final inverse photon slope are seen to be roughly of the same order of magnitude as those from viscous corrections to the photon emission rates (${\sim\,}10\%$ for $T_\mathrm{eff}$), and thus too small to be experimentally resolved with the present experimental accuracy of $T_\mathrm{eff}$. We note that, for both RHIC and LHC energies, the calculated inverse slopes are consistent (within errors) with the experimentally measured values, although near the high end of the observationally allowed band for RHIC.

We conclude that thermal photons can indeed be used as a thermometer in relativistic nuclear collisions, but that their interpretation requires a dynamical model which has the sophistication demanded by the wealth of hadronic data that currently exist at RHIC and at the LHC.  We observe that the large observed effective temperatures of thermal photons emitted from heavy-ion collisions, and their significant increase from RHIC to LHC energies, reflect mostly the strong radial flow generated in these collisions and do not {\it directly} prove the emission of electromagnetic radiation from quark-gluon plasma with temperatures well above $T_\mathrm{c}$. In particular, they are not representative of the initial temperature of the QGP generated in the collision. We hasten to say, however, that a hot and dense early stage of the expanding medium is necessary to generate (either hydrodynamically or by pre-equilibrium evolution) the large radial flow causing the high effective photon temperatures (inverse slopes). The dense early stage thus plays a crucial role, even if it does not dominate the electromagnetic radiation.

Our conclusion that the measured thermal photons are mostly emitted at a relatively late, strongly flowing stage of the fireball is consistent with the unexpectedly \cite{Chatterjee:2005de,vanHees:2011vb,Rapp:2011is,Dion:2011pp,Chatterjee:2013naa,Shen:2013cca} large photon elliptic flow measured by both PHENIX \cite{Adare:2011zr} and ALICE \cite{Lohner:2012ct}. In fact, these data appear to require an even stronger weighting of photon emission towards the end of the expansion stage where flow is strong \cite{vanHees:2011vb,Lohner:2012ct}. Our finding that the experimentally measured centrality dependence of thermal photon yields appears to be closer to what our model predicts for hadronic radiation than for QGP radiation lends further support to this line of thought. Making a compelling argument for photon radiation from the earliest and hottest stages of the fireball requires combining the photon inverse slope measurements with other electromagnetic observables and a detailed and quantitative comparison with theory. To be convincing, the argument must be based on measurements and theories that determine $T_\mathrm{eff}$ for thermal photons with about 5\% precision. While it is unlikely that future improvements in the theoretical rates change their effective temperatures by a large margin (see the small difference between QGP and HG inverse slopes (green line) in Fig.~\ref{F1}), it is possible that the currently used $T$-dependent rates receive corrections that increase photon yields in the critical quark-hadron transition region, and that further improvements in the dynamical modeling, in particular towards the end of the collision where hydrodynamics begins to break down, will change the weighting of the emission rates by altering the space-time volumes corresponding to each temperature slice.

{\bf Acknowledgements:} The authors thank the Yukawa Institute for Theoretical Physics, Kyoto University, where this work was discussed during the YITP-T-13-05 workshop on New Frontiers in QCD. The authors wish to thank Mike Lisa for a very useful suggestion for presenting some of our results. This work was supported in part by the U.S.\ Department of Energy under Grants No.~\rm{DE-SC0004286} and (within the framework of the JET Collaboration) \rm{DE-SC0004104}, and in part by the Natural Sciences and Engineering Research Council of Canada. J.-F.P. acknowledges scholarships from Hydro-Quebec and from FRQNT, and C. G. acknowledges support from the Hessian initiative for excellence (LOEWE) through the Helmholtz International Center for FAIR (HIC for FAIR).





\begin{thebibliography}{99}
\bibitem{pQCD}The minimum photon $p_T$ at which the pQCD calculation can be made is
determined by the choice of factorisation, renormalisation and
fragmentation scales, which are all chosen here to be equal to a constant
times the photon $p_T$. At  RHIC, this constant is chosen to
be $1/2$, yielding good agreement of the pQCD calculation  with low $p_T$ proton-proton data from PHENIX. In the absence of low $p_T$ p+p
data at the LHC, the scales were chosen to be $p_T$. The consequence of this choice is that
the minimum $p_T$ at which the pQCD calculation can be made at RHIC is
3.0 GeV/c, while the minimum at the LHC is 1.5 GeV/c. The fits were done using the calculated results above these values. The reliability of the specific form chosen for the extrapolation can
be verified by setting the photon momentum scaling factor to a larger number (e.g. 2
instead of 1/2 or 1), which permits to
extend the pQCD calculation to lower $p_T$. We have made sure that in
such a case the calculation could still be well-described by a function
of the form $A/(1+p_T/p_0)^n$.
\end{thebibliography}
\end{document}